\documentclass{ws-procs961x669}

\makeatletter
\renewcommand{\ps@plain}{%
  \renewcommand{\@oddhead}{\hfil{\footnotesize%
    A contribution to the Julian Schwinger Centennial Conference, %
    7--12 February 2018, Singapore}\hfil}%
  \renewcommand{\@evenhead}{\@oddhead}%
  \renewcommand{\@oddfoot}{\hfil\thepage}%
  \renewcommand{\@evenfoot}{\thepage\hfil}%
}
\makeatother
\crop[off]

\begin{document}
\title{Julian Schwinger --- Recollections from many decades}

\author{Stanley Deser}

\address{Walter Burke Institute for Theoretical Physics,\\
California Institute of Technology, Pasadena, CA 91125;\\
Physics Department, Brandeis University, Waltham, MA 02454;\\
{deser@brandeis.edu}}

\begin{abstract}
I present some reminiscences, both personal and scientific, over a lifetime of
admiration and friendship with one of the Grandmasters of our subject.
\end{abstract}

\section*{}

Dear students, friends, and admirers of Julian Schwinger, or all three. 
We are here to celebrate and commemorate a century since Julian's birth.
He only lived three quarters of that period, unfortunately dying far too young
at 76, but left us a great legacy.
Being in this Conference's history section, I will try to discuss the   
life and work as I saw it, minus technicalities.

I knew Julian for three-fifths of his life, a reasonable fraction.
It began when I arrived as a graduate student in the fall of 1949.
I didn't know much physics, nor did I know who Julian was, but I was soon
educated on the latter.
In fact, I sat in on three of his quantum mechanics courses, all different.
Like everybody else who wanted to do theory, I was convinced that Julian
should be my mentor.
He was willing to accept just about anybody, but he was chary with his time,
as all his students know.

Since he has saved my life so many times, I feel I should begin by giving some
examples.  
The system at Harvard in my day, if you wanted to do theory, required you to take
a qualifying exam, usually in something called Math and Mechanics, which
covered various sins.
One was supposed to bone up on that during one's second academic year; a jury
of one's would-be advisor plus two other people was then convened.
The day duly came and Julian arrived, flanked by Abe Klein and Bob Karplus,
two up-and-coming assistant professors whose careers depended critically on
sufficiently impressing Julian so they could get good positions elsewhere ---
one didn't get promoted from within.
They had discovered something in their latest calculations, some particularly
uninteresting but technical stuff called dilogarithms, which are now, I
suppose, taught in kindergarten but in those days unknown to anyone ---
certainly to me.   
They proceeded to show Julian how brilliant and clever they were, at my
expense, so that after the first few words, I was totally excluded from
everything, and after an hour and a half of this, they turned to me and
pityingly asked me a question like what two plus two was, at which point I
couldn't even have answered one plus one.
And so this terrible ordeal ended, I walked out, and two minutes later Julian
came and said, ``you realize you failed your qualifying exam,'' and I said
``yes,'' and there was a little pause and he  went on, ``don't worry about
it.''
I think this miracle (and miracle it was --- no one else failed M\&M) may have
been due to my performance on an advanced electrodynamics course I had just
taken with him.

Then I started on my thesis.
I think I probably saw Julian for a total --- just on the upper limit --- of
about ten hours during those two years.
One day, in the spring of my fourth year, I asked Julian, when would I could
possibly think of finishing up.
When he  replied ``right now if you want'' --- this was shortly before the
strict Harvard deadline for submitting a thesis, I was not going to let this
opportunity slide; somehow it all got done and typed on a Bible paper, only
available in one place in the world, and bound in one particular way, and all
the rest of it. 
Although the thesis was mediocre, I was handed my Ph.D. by James Bryant
Conant, in his last year of a long tenure as president of the university.

Rescue number two was a bit more indirect.
In those days, Julian would simply phone Oppenheimer at the Institute for
Advanced Study, tell him who his latest graduates were and Oppenheimer would
take them, no applications or recommendations.
Unfortunately, the year before mine, Julian's choice at that point was a very
strange guy, we'll not name names, who was found in his first year at the
Institute climbing the wall of some estate in Princeton, something frowned
upon at such a rich community.
The whole thing was handled very well, all airbrushed out.
He disappeared, and I'm told became a successful psychoanalyst, but that could
be apocryphal.
In any case, Oppenheimer was taking no chances, so he told Julian that his two
picks, Roger Newton and I, had better show up and pass a psychiatric exam.

In those days Oppenheimer still had his clearance, so two FBI agents were
guarding his files; I walked past with trepidation, but all Oppenheimer did
was ask me what my thesis was about, the title of which I told him.
He immediately told me (a) what was in the thesis and (b) why it was wrong.
He was way off the mark, at least on point (a); he had no idea whatsoever, but
that was in his style.
At least I didn't have any obvious tics.
I was vetted also by the younger permanent people at the Institute, and Roger
also passed with flying colors.
We were installed at the Institute where I had my two years, and not so much
contact with Julian.
However he saved me because when I arrived at the Institute not too sure what
to do, I was immediately pounced on by Murph Goldberger and Walther Thirring
who were both visiting there.
They said, ``you must know all of Julian's tricks, so let's get moving and
apply them to the following project.''
Of course I didn't know Julian's tricks, but it in fact provided my first
successful extra-Ph.D. experience and did use some of them after all.  
I should mention --- going back a bit --- that before you start on a thesis,
you're given a little test problem by your advisor.
Julian gave me the little test problem, of which I had no idea whatsoever at
all what to do, the reason was that this little problem was the beginning of
his celebrated National Academy of Sciences series that to this very day is a
standard tool.
So when he showed me what he had done, I realized why I hadn't a clue as to
what to do.
Well, that too he accepted.
So I learned from that that one should do onto others and give would-be
graduate students a certain amount of leeway, perhaps not as much as he gave
me, but still.

Then came my second postdoc stage.
After the two years at the Institute, I went off for two years to the Niels
Bohr Institute in Denmark, which was a difficult period for me.
I only wrote one paper, which was furthermore wrong, although wrong in an
interesting way.
In any case, in those days especially, I hadn't realized that, once you go
into exile, you no longer exist in the United States, because you're not in
any loop.
Fortunately, Julian came by that summer, visiting Denmark with Clarice, and he
again saved my life by offering me one year as his assistant as an Instructor
at Harvard, while I found my footing back home.
That was truly critical, because being married having a baby, it was clear
that I needed some sort of a job.

He then also recommended me for my first faculty position at Brandeis.
So, this was the support I got from Julian: his faith in me was truly beyond
any requirement.

His greatest confidence in me occurred much later.
I was an invited visitor to UCLA, where Julian had moved, and used to stay in
his house.
Once I came it was during one of those oil embargos when you couldn't get any
gas for your car, especially in California.
Julian  lived in Bel Air, which had, and has still, for all I know, one and
only one gas station, at some chi-chi little shopping center.
It was going to open at 7AM until the gas ran out by 7:30, and Julian was of
course in a terrible quandary because 7AM is too late for staying up and far
too early for getting up.
I was still on Eastern time, so 7AM suited me fine, but would he
entrust his precious sports car?
He agonized all evening and then finally handed me the keys, gave
me a three-hour lecture on how to drive, and I'm sure had a very restless
night.
I arrived at 7AM, surrounded by all the neighborhood Bentleys and
Rolls-Royces, chauffeurs waiting in line, but I did manage to snag sufficient
gas for the next period and avoided having any dents in the car, which Julian
inspected~carefully.

Our relations became more even with time.
In particular, after the birth of supergravity in '76, Julian
asked me to come for a weekend tutorial for him and his entourage at UCLA.
So there, on a Saturday at some ungodly early hour like 10AM, we started on a
full Soviet-style two day session; Julian would say, ``I don't understand what
this is,''  and I replied ``come on, Julian, you invented it all,'' and
reminded him of the Rarita--Schwinger equation, which he did indeed vaguely
remember --- they had actually a fairly ugly form for it --- but they had
found it.
I suggested that in fact Julian should have discovered supergravity, as he had
all the tools.
He was of course one of the few people in the United States, back when he was
student in the '30s, who even knew general relativity.
Just like he knew quantum mechanics.
You all know the famous story of how when he was flunking out of City College,
Lloyd Motz brought him to Rabi to try to get him a transfer to
Columbia; he must have been in the teens. 
He was put on a bench while they argued about some quantum mechanics problem
unsuccessfully, and then he spoke up, adding the one word that explained it all.
Motz and Rabi finally realized he was in the room, and the rest is history. 

He was also an extremely cultured person, and though very shy, he was really
quite up on a number of extra-physical things that one might not have
expected.
I realized that also during the year that I was his assistant in '57--58,
because I was then in the position that we graduate students used to envy.
When we were students, with his assistants, instead of seeing us he would walk
off to the restaurant of his choice, there was only one at Harvard Square
those days that was even semi-edible, and leave us in the lurch.
So this time it was my turn to leave the students in the lurch, and we would
talk about all sorts of things, physics, and non;
a great educational opportunity.

The range of Julian's discoveries and inventions, formal as well as directly
physical, is, as we know, enormous.
He was of course a great master of Green functions and everything related to
gauge fields.
He was one of the predecessors of the weak interaction theories that were then
soon developed --- Shelly Glashow was his student.

Julian was extremely active not only during the great triumph of QED, 
quantum electrodynamics, but after that stayed very much in touch with
developments.
However, and this is my own theory, the Moses complex, that great men who are
handed the truth from up above, are fated never to set foot in the promised
land, and Julian is certainly an example.
He felt at a certain point, especially in his last times at Harvard, that he
was more or less sidetracked from what was then the main line, and that's when
he made his motto ``If you can't join 'em, beat 'em,'' the origin of source
theory, which engaged him for quite a long time, and which of course was a
very interesting way to look at quantum field theory, although not really as
productive as he might have hoped.
He also had his engagements with cold fusion, and I think it was all part of a
reaction --- the Moses reaction, not being able quite to go to the next stage,
such as it is, our standard model ``promised land.''

But he provided an enormous amount of impetus through his students.
I remind you that Julian had something like 72 or so Ph.D.s to his credit.
That is an amazing number, four of whom --- if you count that way --- were
Nobel Prize winners, not bad.
Roy Glauber, Shelly Glashow, Ben Mottelson and Walter Kohn, although he got it
in Chemistry, was very much a Schwinger product.
In fact, my first quantum mechanics course was taught by Walter, it was one of
the many Schwinger QM variants.
So, his influence both with the early Ph.D.s and of course later on with the
cohort at UCLA, many of whom will be speaking and reminiscing here, should be
very much counted as part of his contribution to our subject.

But  really --- when you think about it, there's almost nothing we do
in theory that doesn't somewhere bear Schwinger's imprint.
In fact, in the old days, there was a great form that you could fill out in
order to publish a paper: it  started with ``According to Schwinger \dots,''
then you would put in the equation of your choice, and then went on and said
``\dots\ now using the Green function appropriate to this problem we discover
that \dots,'' and the paper was guaranteed to get into the Physical Review.
That was not so far from true.

But of course, Julian was much more than that.
He was involved in early postwar nuclear physics; I took a full year's nuclear
physics course from him.
This was really dirty nuclear physics and its phenomenology, effective range
theory, scattering and bound states, again from an effective theory point of
view, it was a very powerful tool.
It was not quarks, but it was a way to understand, classify, and normalize
low-energy nuclear physics, as it was then practiced, a forefront field. 

He had an enormous effect on classical electrodynamics.
We all know that during the war his work on waveguides and propagation was
not only very useful to the war effort, but really began a whole field of
investigations in that time --- and then of course, QED as I mentioned.
I still remember his lecturing, his derivation, first of course of
$\alpha/2\pi$, as he did it in class, and then of the Lamb shift, the really
great achievements of QED.
He did it in real time on the blackboard, sans notes.
I have since taught the Lamb shift and I couldn't do any better than Julian
had done 20 years earlier.   

Sometimes, of course, like all great men he faltered.
There was an early famous incident.
Every Wednesday afternoon the  joint theoretical seminar used to oscillate
between Harvard and MIT.
Julian started one, claiming to have completely solved the closed form of QED,
which was clearly going a little too far.
Finally, at some point Francis Low pointed out that in order to do this Julian
had assumed that a four-point function was simply the product of two two-point
functions, and of course if you assume that, you're talking about a free
theory.
So that particular attempt did not work.
I mention this to say that Julian, like all great men, was of course fallible,
but fallible in a very trivial way, never on the real essentials.
I think the only person who was infallible was Enrico Fermi; he was called the
pope because of that.
But in Julian's frontier explorations, you had to take the risk of being wrong
in order to get anywhere.
He also was interested in general relativity later, so when we --- ADM:
Arnowitt, Misner, and I --- especially Arnowitt and I at the end of my year at
Harvard began our work on the theory, Julian got interested and wrote
a couple of papers.

He also tried to do something about supersymmetry which didn't quite work out.
There is actually very little in modern theory that he was not at least
cognizant of if not actively pursuing, so that the brain never really stopped
working.
For example, he set Wally Gilbert, George Sudarshan and me on dispersion work
during that Harvard year that is still useful. 

As I said, later on he became a little bit isolated from the mainstream
community, but that's not that he was unaware of what they were doing, he knew
what was going on, and although he never made it into the standard model, he
certainly laid its groundwork in many directions.
The work that he did in electrodynamics evolved into the heroic calculations
that his graduate students --- Charlie Sommerfield, for example --- utilized
to do the two-loop magnetic anomalous moment corrections.   
That Julian  stayed away from Feynman diagrams is a trivial difference which
has wrongly been blown up.
Whether you have a Green function or a line makes no difference; at the end of
the day, you are doing the same integrals.
I give these --- among many other --- examples to indicate the sheer scope of
Julian's inventions.
In the early days, the Rarita--Schwinger equation, which was the first serious
attempt at going beyond spin~1 --- was completely virgin ground at that point,
and of course the significance of going beyond spin~1 and its problems did not
emerge until very much later, again very much before its time as often was
with Julian and his work. 

During the war, he, like all the other physicists, concentrated on
applications --- all were either in Los Alamos or the Rad Lab, where Julian
chose to go.
It was not until after the war and he was lured to Harvard that he was able to
really go full speed again in fundamental physics. 
The aura that Julian had around him at Harvard was really unparalleled and
well-earned.
It was a sad day for Harvard when he was lured to UCLA, although they did
quite well with Steven Weinberg as a replacement, I should immediately add.
UCLA was greatly enlivened by the presence of Julian and his group, and
although I suspect that eternal sunshine and tennis had something to do with
his move, he was really quite happy at UCLA.   

He died of a cancer with a guaranteed lethal outcome, but I think he enjoyed
his life to the very end.
Clarice was his ideal companion.
He has left the memories we all know for his successors.
He has made us all his ex-students.
Even people who were not yet alive when he died benefited from the foundations
that he laid to our field.
I can think of no better exemplar in every respect.
It was, for me, a great honor to be his student, as I'm sure will be echoed
by everyone in this room. 

Thank you.
 
\section*{Aknowledgments}
This work was supported by grants NSF PHY-1266107 and DOE\#desc0011632. It
is a transcription of an invited video contribution to the Schwinger 100
Symposium, Singapore 2018.

\end{document}